Identifying Covariational Reasoning Behaviors in Expert Physicists in Graphing Tasks

Charlotte Zimmerman
University of Washington

Alexis Olsho
University of Washington

Michael Loverude
California State University Fullerton

Suzanne White Brahmia
University of Washington

*Covariational reasoning — how one thinks about the way changes in one quantity affect another quantity — is essential to calculus and physics instruction alike. As physics is often centered on understanding and predicting changes in quantities, it is an excellent discipline to develop covariational reasoning. However, while significant work has been done on covariational reasoning in mathematics education research, it is only beginning to be studied in physics contexts. This work presents preliminary results from an investigation into expert physicists' covariational reasoning in a replication study of Hobson and Moore's 2017 investigation of covariational reasoning modes in mathematics graduate students. Additionally, we expand on this work to include results from a study that uses slightly more complex physics-context questions. Two behavioral modes were identified across contexts that appear distinct from those articulated in the Hobson and Moore study: the use of compiled relationships and neighborhood analysis.*

**Introduction**

The ability to understand quantities and how they relate to each other is an important skill in day-to-day life. We expect, as educators in mathematics-based disciplines, that our students leave our classrooms with this *quantitative literacy* and carry it forward into the world. Covariational reasoning – how small changes in one quantity affect another quantity – is an essential piece of quantitative literacy. Covariational reasoning has been studied significantly in mathematics education research, but is only beginning to be examined in the physics education realm. Physics, where the focus is often on understanding and quantifying change in a physical system, is an important and useful setting for developing students' covariational reasoning. Therefore, it is integral that we better understand how covariational reasoning manifests in physics so as to more directly enhance our students' quantitative literacy across contexts.

In prior research, covariational reasoning has been shown to be an essential part of student understanding in precalculus and calculus courses (Thompson, 1994; Saldanha and Thompson, 1998; Oehrtman, Carlson, & Thompson, 2008), and plays a key role in forming conceptual understanding of derivatives, integration, and graphing functions (Johnson, 2015; Moore, Paoletti, & Musgrave, 2013; Castillo-Garsow, Johnson, & Moore, 2013; Paoletti & Moore, 2017; Carlson, 1998). Carlson, Jacobs, Coe, Larsen, and Hsu (2002) developed a framework to describe the mental actions of students engaged in covariational reasoning. Here, *mental action* is used to describe a specific behavior in students' reasoning. In 2017, Hobson and Moore reported the mental actions and associated reasoning modes of expert mathematicians (defined in their paper as mathematics graduate students with teaching experience at the introductory level) when engaged in graphing tasks designed to elicit covariational reasoning. After reviewing these works, we were interested how covariational reasoning manifests in expert physicists' reasoning.

We seek to investigate two questions: (a) how do expert physicist behaviors compare to those found by Hobson and Moore (2017) in expert mathematicians, and (b) are these behaviors consistent outside motion-based graphing tasks? The first follows from our literature review, and the second arises because the original graphing tasks are focused on motion-based systems. Producing motion graphs is a heavily practiced skill for physics graduate students engaged in teaching introductory physics, and we were curious if the behaviors we saw were simply due to the high level of familiarity with the types of questions asked. Therefore, we performed two rounds of think-aloud interviews with physics graduate students, following the experimental model provided by Hobson and Moore's work. We claim that physics experts employ two behaviors that are notably different from those articulated in Hobson and Moore's 2017 paper; we call these *using compiled relationships* and *neighborhood analysis*.

## Methods

We interviewed two rounds of ten physics graduate students, five of whom participated in both interview rounds. We chose to study graduate students since they are considered to have expert-level knowledge for our introductory-level questions. Each question was designed with an accompanying animation designed to mimic the method used by Hobson and Moore (2017). For both rounds of interviews, participants were given a computer with animations that represented the problems and a paper sheet with the question prompts. They were able to play the animations at will and move between questions until they felt satisfied with their answers.

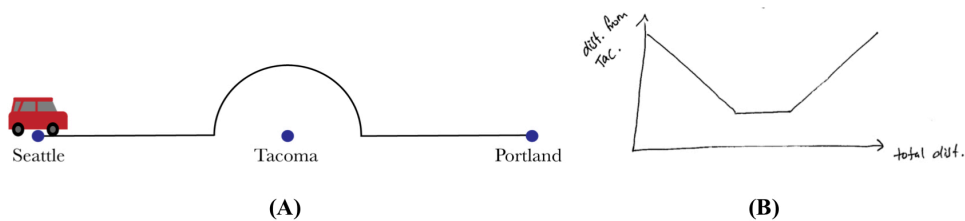

*Figure 1: (a) An animation still, and (b) an example of participant work from "Going Around Tacoma". Participants were given the following prompt: "A few students are going on a road trip, and decide to drive from Seattle to Portland. They want to avoid the traffic around Tacoma, and take the path shown in the animation. Draw a graph of the students' distance from Tacoma vs. their total distance traveled in the space below."*

The animations and text from the first set of interviews were adapted from Hobson and Moore (2017) in an effort to reproduce their study with physics graduate students. The items are based around relating two distances in a situation regarding a moving object. For example, our version of Hobson and Moore's *Going Around Gainesville* task, which we call *Going Around Tacoma*, asks the participant to relate the distance from Tacoma to the total distance traveled as a car drives from Seattle to Portland (Figure 1). The second round of interviews was designed in a similar format, but with non-motion based physics contexts. For example, in *Rigamarole*, the participant is asked to relate the gravitational potential to the total distance traveled as a spaceship moves around a planet (Figure 2).

Audio recordings of the interviews were collected, along with the participants' written work. Initial transcripts were produced using the computer program Otter, and subsequently hand corrected (Otter, 2019). The transcripts were coded by one researcher using a modified approach to grounded theory, as the coding was informed by the researcher's literature review. Inter-reliability was confirmed by the other authors in the study.

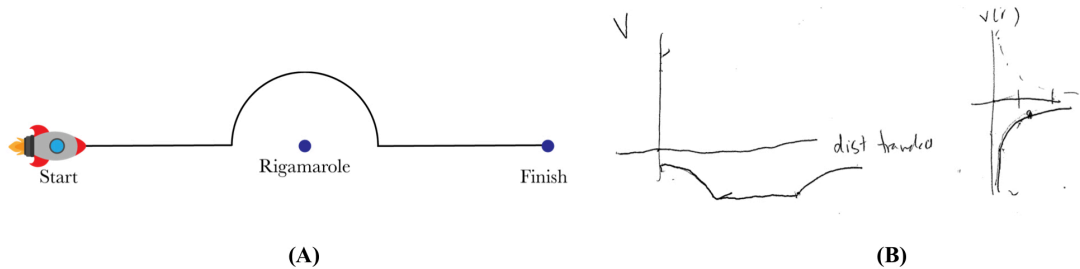

*Figure 2: (a) An animation still, and (b) an example of participant work from "Rigamarole". Participants were given the following prompt: "Spencer is piloting a spaceship near the planet Rigamarole. Not wanting to crash into the surface, he carefully directs the ship around the planet, and continues on his journey. The animation represents a simplification of his journey. Create a graph that relates the total distance traveled by Spencer's ship and the gravitational potential energy of the entire system (Spencer's ship and the planet Rigamarole)."*

## Results

The transcripts from both rounds of interviews yielded rich, detailed reasoning patterns about covariation and therefore analysis is ongoing. In preliminary results, we have identified two reasoning patterns to be discussed here: using compiled relationships and neighborhood analysis. *Using compiled relationships* describes when participants use a previously known relationship between quantities to draw the graph rather than considering the problem at hand directly. *Neighborhood analysis* refers to when participants draw their graphs by identifying physically significant points, considering the slope of the graph around those points (in the points' "neighborhood"), and then connecting those slopes to form a smooth curve.

### Using Compiled Relationships

In introductory physics, there exists a small, finite number of functions that are used to model physical situations. We speculate these are drawn upon by expert physicists and employed to reduce cognitive load. Here we discuss two examples of compiled relationships being used: a constant circular motion model employed to assign trigonometric functions in *Ferris Wheel* (Figure 3) and using the inverse relationship between potential and radius from the source object ($V \propto 1/R$) to generate the graph for *Rigamarole*.

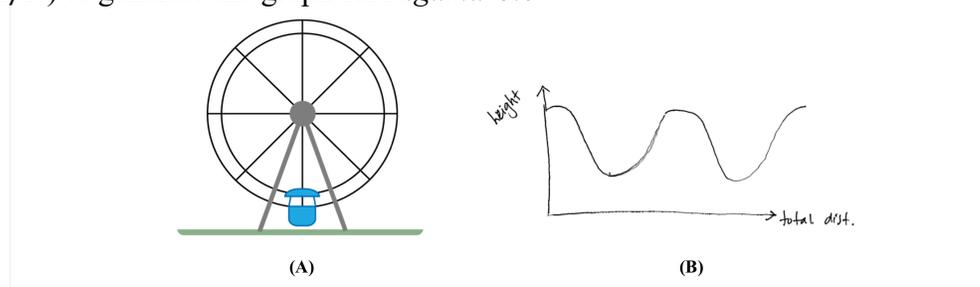

*Figure 3: (a) An animation still, and (b) an example of participant work from "Ferris Wheel". Participants were given the following prompt: "As shown in the animation, a cart moves around a Ferris Wheel. Draw the graph of the height of the cart from the ground vs its total distance travelled in the space below."*

*Ferris Wheel* was given in the first round of interviews. In our analysis we saw compiled relationships used when participants applied a uniform circular motion physics model, ubiquitous in introductory physics. Immediately, one participant stated: "I feel like this is where, like, my understanding of trig functions really comes in handy. Because I know this is a circle. And so, the height goes like a trig function." They quickly identified that circles indicate a trigonometric system, and spent their time instead determining the initial conditions rather than the form of the graph itself: "I know it's like, basically a sine wave. But then there's a choice of how to phase it." After deciding that the animation starts with the cart at the top of the curve, and therefore the highest point, they arrived at the correct graph as shown in Figure 3B. Only after significant probing from the interviewer did the student produce reasoning that directly compared the height of the cart from the ground and the distance the cart traveled. In comparison, Hobson and Moore report their participants dividing the Ferris wheel into equal sections and considering the curvature of the graph for each section (2017).

Similarly, in the second round of interviews, *Rigamarole* seemed to spark the use of compiled relationships (Figure 2). For example, one participant used the established physics relationship between potential and position from the source, $V \propto 1/R$, to arrive at their graph (Figure 2B). They began by noticing that position from the source and distance traveled are not the same quantity. They drew a separate graph of the potential with respect to the position, R, from Rigamarole off to the side of the page, and then compared points to determine the shape of the curve:

> "So as I go from distance traveled, I'm basically starting at a fixed $R_o$ and I'm going to a smaller R there [gestures to the start and end points of the first section of Rigamarole]. So V is going to go down… so I can copy and paste this section [referring to the potential plot] to either end of the semi-circle."

Here, the student uses their prior knowledge of how potential behaves with respect to position from a source, and then compares distance traveled and position to arrive at the answer rather than directly considering how the potential changes with respect to distance traveled.

**Neighborhood Analysis**

*Neighborhood analysis* refers to choosing notable points in the problem, analyzing the rate of change between two quantities in the "neighborhood" near each point, and connecting those rates to form a smooth curve that represents the entire scenario. We found that when unable to relate a compiled relationship to the quantities in question, expert physicists often turned to neighborhood analysis as a way of solving the problem. We present two examples here, one from an expert analyzing *Ferris Wheel* that did not immediately see the uniform circular motion relationship, and another solving *Drone* from the second round of interviews (Figure 4).

While the majority of the physics experts interviewed did apply a uniform circular motion model to *Ferris Wheel*, there were a few that instead performed neighborhood analysis. One representative participant began by identifying significant points on the trajectory (at the top, bottom, and sides) and discussing how the height changes near those points. For example, when analyzing the point at the left side of the Ferris wheel, they stated: "The component of the velocity vector that's pointing towards the ground is going faster, which means its height from the ground will be changing faster [than at the top and bottom] at this point." The graduate student used vectors to assist their understanding of how the cart is moving, and arrived at the conclusion that the cart quickly changes height on the sides and slowly changes height at the top and bottom of the Ferris wheel: "So if I want it to be slow, fast, slow… I think it looks like… [draws a sinusoidal-shaped curve]." They arrived at the same-shaped curve as their peers that applied a compiled relationship reasoning pattern, but did so by only examining the covariational relationship at a few points during the cart's journey. In contrast, Hobson and Moore described the students they interviewed as dividing up the entire journey into equal parts and comparing the change between sections. We speculate that by only analyzing the covariation at points that appear *physically* significant, physics experts are reducing their cognitive load once more, perhaps towards finding a compiled relationship.

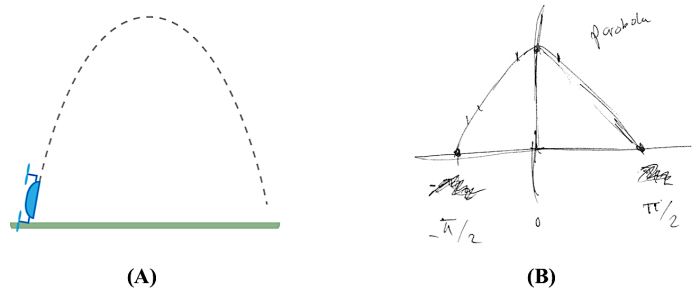

(A)                                    (B)

*Figure 4: (a) An animation still, and (b) an example of participant work from "Drone". Participants were given the following prompt: "A few physics students are playing with a drone, specially designed to reduce air resistance by aligning its axis along the direction of velocity. As the drone moves through the air, the students record the angle of the drone's axis relative to the ground. The animation represents a simplification of the drone's journey. Create a graph that relates the height of the drone from the ground and the angle of the drone's axis with respect to the ground. You may assume the drone experiences negligible air resistance."*

In the second round of interviews, we continued to see participants exhibit neighborhood analysis. One participant in particular wrestled with their interpretation of *Drone* for some time. When they first approached the problem, they had difficulty thinking through the relationship and quickly drew the solution shown, articulating little supporting reasoning (Figure 4B). Unprompted, the student later reflected on their answer, remarking, "I guess it makes sense… because the angle is changing really quickly near the middle… if I look at two points in the middle, like, it's changing faster than, say, two points here [gestures to the start of the trajectory]." While they have difficulty conceptualizing how the two quantities were related upon reading the problem, the student is ultimately able to develop an explanation for how the quantities are changing with respect to each other by comparing two representative neighborhoods rather than equal sections of change.

## Conclusion

Using compiled relationships and neighborhood analysis are distinct behaviors we identified expert physicists exhibit during interviews. This suggests that physics experts are thinking about covariation differently at times compared to mathematics experts. Physics has recently been discussed in physics education research as an conceptual blend of physics concepts and mathematical understanding. It is valuable to understand how instructors of mathematics and those of physics are using and modeling these behaviors to their students (Eichenlaub and Redish, 2019; Brahmia, Boudreaux and Kanim, 2016). The difference in covariational reasoning we observed suggests that further work is required to better understand how covariational reasoning is used in physics as it likely has an impact on student understanding across disciplines.

## Acknowledgements

We thank Natalie Hobson and Kevin C. Moore for their permission for our replication of their work with physics graduate students, and their interest in our continuation of this project. This work was made possible by funding from the University of Washington and by the National Science Foundation IUSE grants: DUE-1832880, DUE-1832836, and DUE-1833050.